\documentclass[useAMS,usenatbib]{mn2e}
\usepackage{graphicx}

\def\ltsima{$\; \buildrel < \over \sim \;$}
\def\simlt{\lower.5ex\hbox{\ltsima}}
\def\gtsima{$\; \buildrel > \over \sim \;$}
\def\simgt{\lower.5ex\hbox{\gtsima}}

\title[Pal 5 stream substructures]{Exploring the reality of density substructures in the Palomar 5 stellar stream}
\author[G. F. Thomas, et al]{Guillaume F.~Thomas,$^{1}$\thanks{E-mail:
guillaume.thomas@astro.unistra.fr}
R.~Ibata,$^{1}$ B.~Famaey,$^{1}$ N. F.~Martin,$^{1}$ G. F. Lewis$^{2}$ \\
$^{1}$Observatoire astronomique de Strasbourg, Universit\'e de Strasbourg, CNRS, UMR 7550, 11 rue de l'Universit\'e, F-67000 Strasbourg, France\\
$^{2}$Sydney Institute for Astronomy, School of Physics, A28, The University of Sydney, NSW 2006, Australia
}

\begin{document}

\date{Accepted 2016 May 15. Received 2016 May 15; in original form 2016 March 17}

\pagerange{\pageref{firstpage}--\pageref{lastpage}} \pubyear{2016}

\maketitle

\label{firstpage}

\begin{abstract}
We present an analysis of the presence of substructures in the stellar stream of the Palomar 5 globular cluster, as derived from Sloan Digital Sky Survey data. Using a matched filter technique, we recover the positions and sizes of overdensities reported in previous studies. To explore the reality of these structures, we also create an artificial model of the stream, in which we construct a realistic background on top of which we add a perfectly smooth stream structure, taking into account the effects of photometric completeness and interstellar extinction. We find that the smooth artificial stream then shows similarly-pronounced substructures as the real structure. Interestingly, our best-fit N-body simulation does display real projected density variations linked to stellar epicyclic motions, but these become less significant when taking into account the SDSS star-count constraints. The substructures found when applying our matched filter technique to the N-body particles converted into observable stars are thus mostly unrelated to these epicyclic motions. This analysis suggests that the majority of the previously-detected substructures along the tidal tail of Palomar 5 are artefacts of observational inhomogeneities.
\end{abstract}


\section{Introduction}

The stellar streams seen around the giant galaxies in the Local Volume \citep{odenkirchen_2001,grillmair_2006a,grillmair_2006b,belokurov_2006,grillmair_2009,ibata_2001,ibata_2014b,martinez-delgado_2010} are the consequence of the disruption by tidal forces of satellites that orbit around them, whether dwarf galaxies or globular clusters. These streams are particularly interesting probes of the global shape of the gravitational potential \citep[e.g.,][]{varghese_2011,sanders_2013} in an environment which is less perturbed and complex than galactic disks \citep{monari_2016}. The detailed inner structure of streams, and in particular their structural over- and underdensities can, in principle, constrain the granularity of the potential and the abundance of dark matter subhalos \citep{ibata_2002,johnston_2002,yoon_2011,carlberg_2012,ngan_2014}.

In this study, we focus on the tidal tails that are escaping from the globular cluster Palomar 5 (hereafter Pal 5), observed for the first time by \cite{odenkirchen_2001} in the Sloan Digital Sky Survey (SDSS) commissioning data, who estimated a length of $\sim 2.6\degr$ for this structure. However, this first detection of the stream was limited by the boundary of the SDSS commissioning data footprint and subsequent SDSS data releases helped to reveal that the stream was substantially longer \citep{rockosi_2002,odenkirchen_2003,grillmair_2006}. With the SDSS DR 4, \cite{grillmair_2006} found that the Pal 5 stream covers at least $22 \degr$, with $18.5 \degr$ in the trailing arm and $3.5 \degr$ in the leading arm.

From these observations, significant variations in the density of stars along the stream were noticed \citep{grillmair_2006,odenkirchen_2009,carlberg_2012}, which could be explained by different physical processes. The N-body simulations of \cite{dehnen_2004} showed that the growth of the Pal 5 stellar stream is a consequence of repeated violent shocks with the disk. Nevertheless, the high frequency of these shocks (approximately every $300$~Myr) reduces the formation of overdensities. In their suite of papers,  \citet{kupper_2008,kupper_2010,kupper_2012} demonstrated that a tidal stream escaping from a globular cluster can be distorted by epicyclic motion. Especially when the cluster is close to its apocenter, like Pal 5, the stream is decomposed into multiple tails which once projected on the sky should be seen as overdensities.

These substructures can also be the scars of dark matter subhalos  crossing the stream \citep{ibata_2002,carlberg_2012a}. However, these encounters heat the stream and fan its ends as illustrated by the study of \cite{ngan_2014} based on the Via Lactea II cosmological simulation. This effect is clearly not seen in the case for the the Pal 5 stream, which is thin and coherent all along its length.

However, the recent observations of \cite{ibata_2016} performed with the Canada-France-Hawaii Telescope (CFHT), which are $\sim 2$ magnitudes deeper than the SDSS, showed that although the density of stars decreases slowly with distance along the Palomar 5 stellar stream, only a single significant overdensity was found. Hence, this raises the question whether the majority of the SDSS substructures are physical in origin, or if they are the consequence of the inhomogeneity of the SDSS and of the small fraction of stars from the stellar stream that it is able to detect. 
    
In this article we will study the detection of overdensities along the Pal 5 stream in the SDSS by comparing their sizes and their positions to smooth models and models derived from N-body simulations. Section \ref{obs} will present the observational data used in our work. Section \ref{method} will explain the extraction method of overdensities (Section \ref{def_method}) and the procedure to create a ``background SDSS'' (Section \ref{background}) and a smooth stream (Section \ref{smooth _stream}); the construction of the N-body simulation will be detailed in Section \ref{Nbody}. The analysis of the overdensities in the observations and in the different modelling methods will be presented in the Section \ref{result}, and finally we will discuss the implications of these results and draw our conclusions in Section \ref{conclusion}.

\section{Observationnal catalogues} \label{obs}

\subsection{SDSS data}
The primary source of observational data for this study is the SDSS DR9. We selected stars with dereddened magnitudes in the \textit{g}-band brighter than $22.5$ and that also have a detection in the \textit{i} and \textit{r}-bands. 

\begin{figure}
  \includegraphics[angle=0, viewport= 1 1 575 283, clip,scale=0.42]{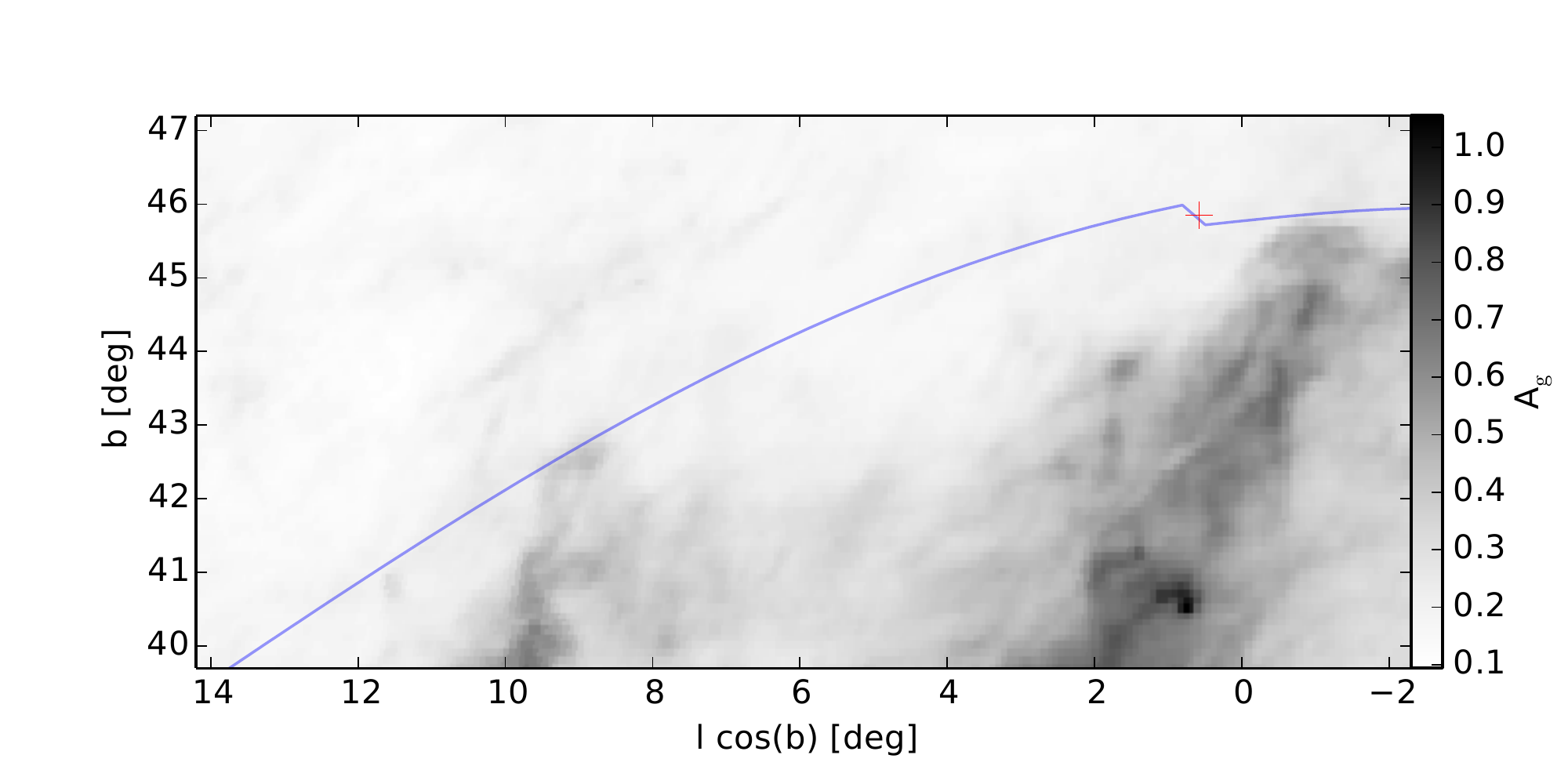}
  \caption{ Extinction map in the g-band derived from \citet{schlegel_1998}. The blue line represents the polynomial fit of the location of the stream from \citet{ibata_2016}. }
\label{extinction}
\end{figure}

Since the foreground extinction varies widely in this region, we correct the magnitude of stars by dereddening them using the extinction map values $E(B-V)$ of \cite{schlegel_1998} and assuming the following conversion coefficients: $A_g/E(B-V) = 3.303$, $A_r/E(B-V) = 2.285$ and $A_i/E(B-V) = 1.698$ \citep{schlafly_2011}. Figure \ref{extinction} shows the extinction in the \textit{g}-band from \cite{schlegel_1998}. 

\begin{figure}
  \includegraphics[angle=0, viewport= 1 1 575 283, clip,scale=0.42]{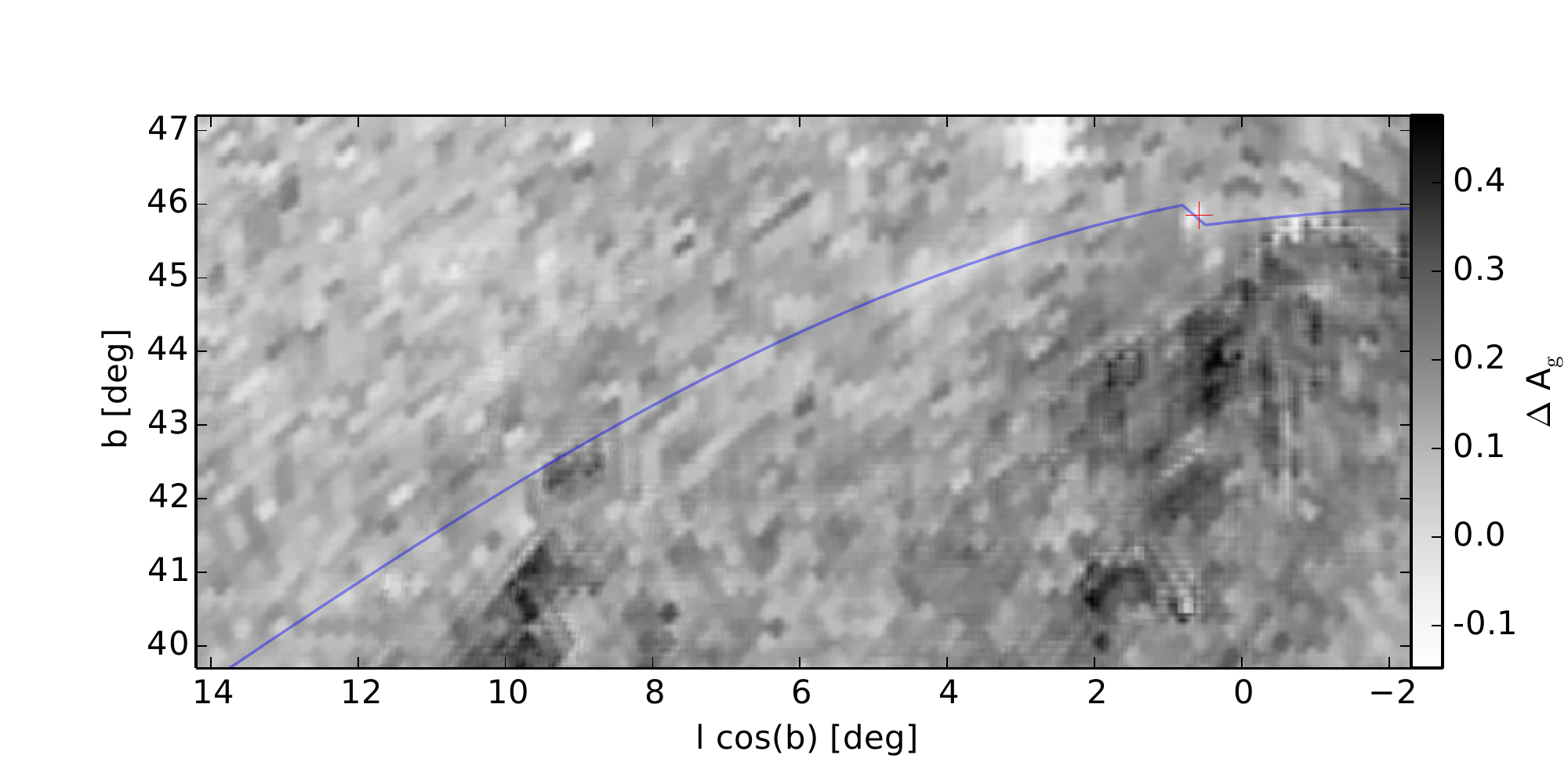}
  \caption{ Difference of extinction in the g-band between the maps of \citet{schlafly_2014} and \citet{schlegel_1998}.}
\label{diff_extinction}
\end{figure}

It is worth noting at this point that the extinction values in this region differ substantially between the \cite{schlafly_2014} and the \cite{schlegel_1998} maps, as we show in Figure \ref{diff_extinction}. Since part of the aim of this paper is to provide a probable explanation for the detections of substructure in the Pal 5 stream derived from SDSS data over the last decade, we choose to use primarily the extinction map of \cite{schlegel_1998}.

\subsection{CFHT data}\label{CFHT}
We also use the 30 Megacam fields ($1\degr \times 1\degr$) obtained by \cite{ibata_2016} at the Canada France Hawaii Telescope (CFHT) in the \textit{g} and \textit{r}-bands. These fields  provide excellent data down to $g=24$, which is significantly deeper than the SDSS limit of $g=22.2$ \citep{ahn_2012}.  

Before dereddeding the stars, we used the following colour equations to correct for the difference between the SDSS and CFHT/MegaCam filters \citep{Regnault_2009}: 

\begin{equation}
 \left.
  \begin{array}{ l }
g_{SDSS}=g_{CFHT} +0.195 \, (g_{CFHT} - r_{CFHT})\\

r_{SDSS}=r_{CFHT} +0.011 \, (g_{CFHT} - r_{CFHT}) \, .
  \end{array}
  \right.
 \label{conversion_mag}
\end{equation}

\begin{figure}
  \includegraphics[angle=0, viewport= 1 1 575 430, clip,scale=0.45]{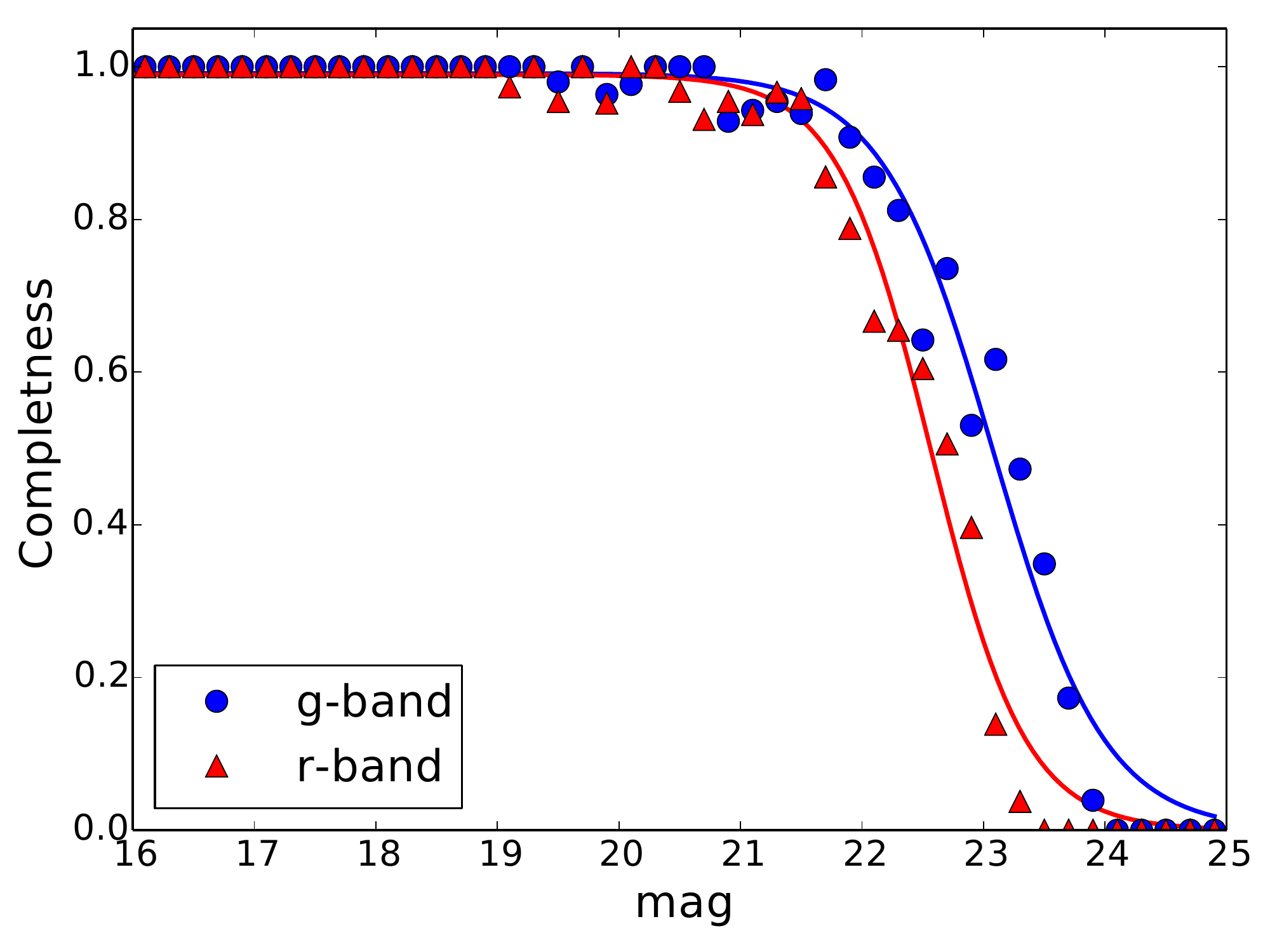}
  \caption{ Completeness of the SDSS survey in an area of 1 deg$^2$ of the tailing arm centred at $(l\, \cos(b)$ , $b) = (3.6,45.2)$, assuming that all stars brighter than 25 in \textit{g} and \textit{r} bands are present in the CFHT fields. }
\label{comp}
\end{figure}

Stars from the globular cluster Pal 5 are generally described as a Single Stellar Population (SSP) with an age of $11.5$ Gyr and a metallicity of $[$Fe/H$]= -1.3$ \citep{smith_2002}.
Since the CFHT data are substantially deeper than the SDSS, it is possible to determine the completeness of the SDSS in the regions around the Pal 5 stream, by assuming that the CFHT is perfectly complete to $25$th magnitude in the \textit{g} and \textit{r} bands.
In both catalogues, we selected in a Colour-Magnitude-Diagram (CMD) the stars close to the Dartmouth isochrone \citep{dotter_2008} with the age and the metallicity cited previously to take into account only the stars that can plausibly be associated with the steam.    
Figure \ref{comp} illustrates the completeness in the \textit{g} and \textit{r} bands in an area of 1 deg$^2$ in the trailing arm. We fitted the following exponential function to these values, where $C_{g,r}$ is the completeness and $g,r$ are the apparent magnitudes.

  \begin{equation}
 \left.
  \begin{array}{ l }
g$-band : $ C_g = 0.99 /(1.0+ \exp( (g-23.08)/0.46) )\\

r$-band : $ C_r = 0.99 /(1.0+ \exp( (r-22.57)/0.39) ) \, .
  \end{array}
  \right.
\end{equation}

The above fits assume the completeness to be 100\% until $21.8$ in the \rm{g}-band and until $21.4$ in the \rm{r}-band, after which the functions drop rapidly. It is important to note that the completeness was calculated on non-dereddened stars. 

This completeness is used below in the construction of the model of a smooth stream (detailed in Section \ref{smooth _stream}) and in the selection of stellar particles form the N-body simulation (detailed in Section \ref{Nbody}).

\section{Method} \label{method}

\subsection{Overdensity detection} \label{def_method}

\begin{figure}
  \includegraphics[angle=0, viewport= 1 1 575 430, clip,scale=0.45]{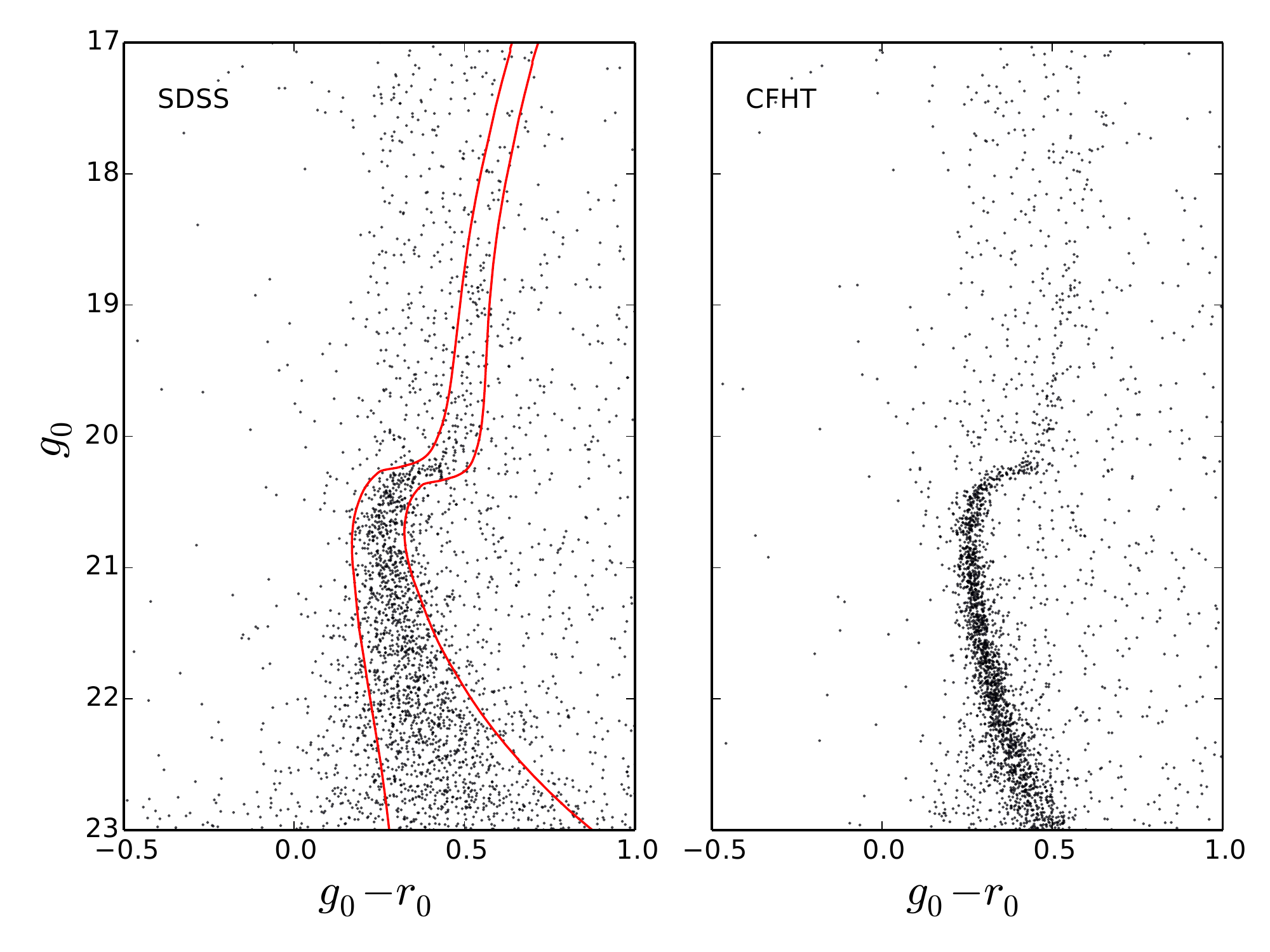}
  \caption{ Photometry in an area of 0.3 deg radius around the center of the globular cluster. The left panel is obtained with the SDSS while the right panel is from our CFHT data. In both panels, the main sequence turnoff of the cluster is clearly visible. Cluster stars are selected from between the red lines, those that are beyond this limit are assumed to be contamination from the background. }
\label{cmd}
\end{figure}

\begin{figure*}
	\includegraphics[angle=0, viewport= 10 10 565 263, clip, width=15cm]{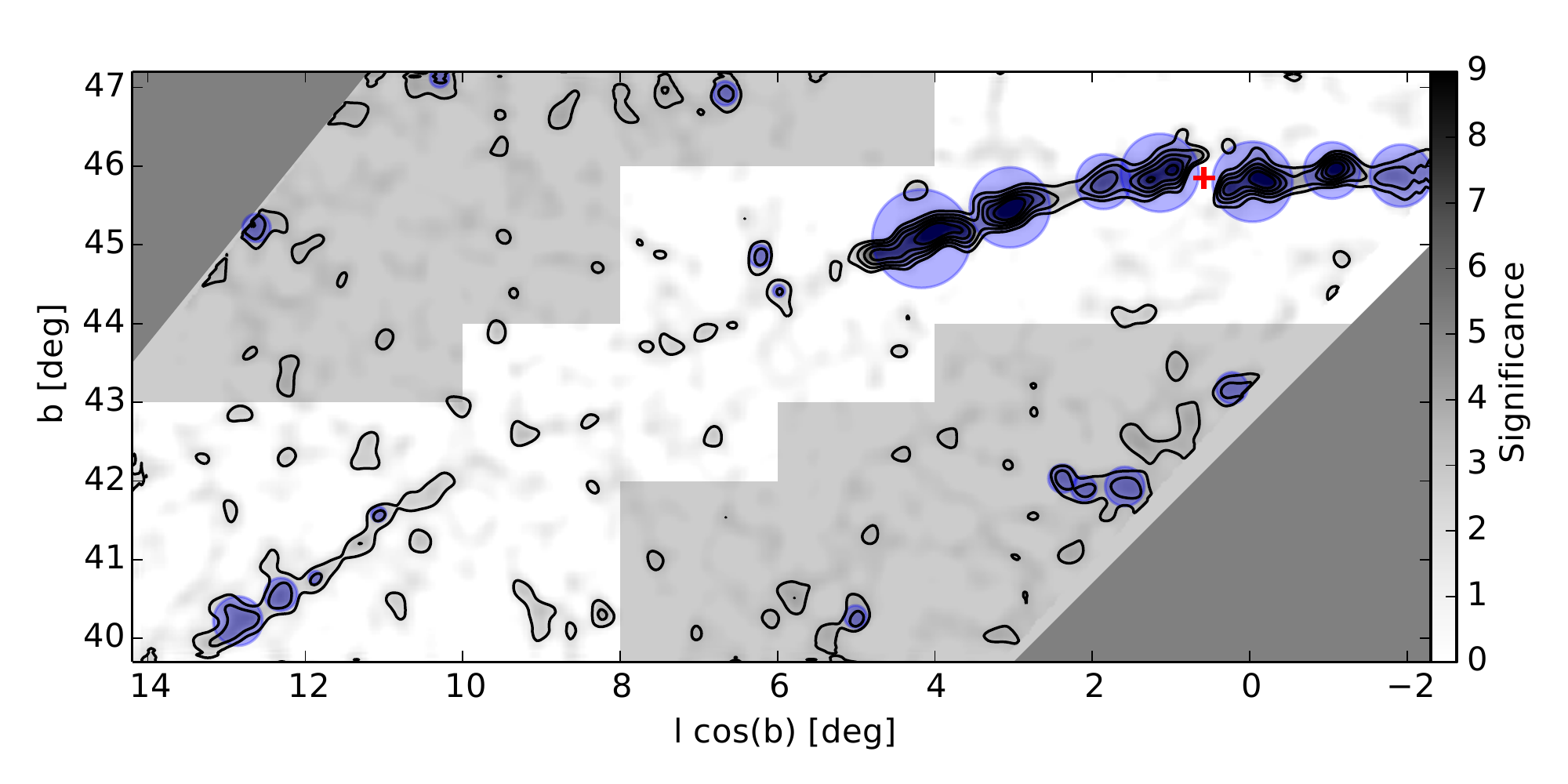}
  \caption{Significance map of the region around Pal 5 in the SDSS survey. The red cross shows the position of the cluster and shaded blue circles are the FWHM of the overdensities. Contours represent significance levels of (2,3.5,5,...). The light grey regions on the border of the footprint are those used to fit the spatial distribution of background stars in the matched filter. }
\label{sig_SDSS}
\end{figure*}

To detect the position and the size of the overdensities of the Pal 5 stream in the SDSS, we apply a very similar method to that used by \cite{kupper_2015}. First, to bring out the stars that could belong to the Pal 5 stream, we select the stars from the SDSS DR9 with dereddened $g_0 < 22.5$ and that follow the colour criterion of \cite{odenkirchen_2003}. We then derive a matched-filter map, which is based on the recipe outlined in \cite{balbinot_2011}. Since the distance variation of the stream is negligible \citep{ibata_2016}, the CMD of the stars along the stream does not change significantly. Hence, the probability function of stars from the stream in the CMD is calculated by reference to the distribution of stars in a radius of $0.3 \degr$ around the center of the cluster (see Figure \ref{cmd}), where we reject the stars that are clearly not associated with the cluster. Since the Pal 5 stream is very thin, we construct the spatial distribution of the background ($\gamma_{bg}$ in \citealt{balbinot_2011}) by fitting a double Legendre polynomial on the borders of the SDSS footprint, represented by the grey regions in Figure \ref{sig_SDSS}, since these regions have small chance to be contaminated by the tidal tails. However, in contrast to \cite{balbinot_2011}, we assume that the CMD of the background does not change over the region of the sky inhabited by the Pal 5 stream. This assumption is reasonable given that the stream lies at high Galactic latitude, and is oriented parallel to the Galactic plane. Furthermore the analysis by \citet{ibata_2016} showed that the ratio of the number of ``background'' stars with magnitudes $18<g_0<19.5$ to those with magnitudes $19.5<g_0<22.0$ remains constant over this region, which supports the assumption that the CMD of the ``background'' does not vary significantly.

Subsequently, to detect the overdensities and their features, we subtract the background component convolved with a Gaussian of $\sigma_2=0.9 \degr$ width, from the ``density map'' matched-filter map convolved with a smaller Gaussian with $\sigma_1=0.115 \degr$.

We calculate the significance in each pixel, $S$, using the formulation of \cite{Koposov_2008}, where $\Sigma$ is the value of the matched-filter in this pixel and $\mathcal{N}$ is the Normal distribution function :

\begin{equation}
S= \sqrt{4 \pi} \sigma_1 \frac{\Sigma (\mathcal{N}(\sigma_1) - \mathcal{N}(\sigma_2) )}{\sqrt{\Sigma \mathcal{N}(\sigma_2)}} \, .
\end{equation}  

To prevent edge artefacts, we do not compute the significance value for pixels within $\sim 1.0 \degr$ of the borders of the SDSS footprint. 

Following \cite{kupper_2015}, we search for large overdensities with the SExtractor algorithm \citep{bertin_1996}, configured to search for groups of at least 10 pixels with a significance larger than 3. The results of this method applied on the SDSS are presented in Figure \ref{sig_SDSS}, where the position and the size of the overdensities are represented with blue circles. Their position, size (FWHM given by SExtractor) and significance are listed in Table \ref{table_SDSS}. They are similar to those of \citet{kupper_2015}, even though some small differences are apparent which are due to the selection box in the CMD and the spatial distribution of the background stars that are not exactly the same in both studies.

\begin{table}
 \centering
  \caption{Positions and sizes of the overdensities in SDSS.}
  \label{table_SDSS}
  \begin{tabular}{@{}lccr@{}}
  \hline
  
   l cos(\textit{b}) (deg) & \textit{b} (deg) & FWHM (deg) & Significance \\
    \hline
  +12.86 & 40.22 & 0.62 & 4.54\\
  +12.62 & 45.22 & 0.36 & 5.06\\
  +12.31 & 40.56 & 0.42 & 4.69\\
  +11.88 & 40.77 & 0.16 & 3.71\\
  +11.07 & 41.58 & 0.19 & 3.84\\
  +10.29 & 47.12 & 0.25 & 3.88\\
  +6.66 & 46.93 & 0.31 & 4.17\\
  +6.22 & 44.86 & 0.28 & 4.31\\
  +5.98 & 44.42 & 0.15 & 3.65\\
  +5.01 & 40.28 & 0.28 & 4.3\\
  +4.17 & 45.08 & 1.25 & 10.67\\
  +3.05 & 45.48 & 1.02 & 11.26\\
  +2.38 & 42.03 & 0.36 & 4.66\\
  +2.11 & 41.9 & 0.32 & 4.13\\
  +1.86 & 45.8 & 0.7 & 6.55\\
  +1.59 & 41.93 & 0.5 & 4.6\\
  +1.14 & 45.92 & 0.99 & 8.73\\
  +0.23 & 43.18 & 0.4 & 4.05\\
  -0.04 & 45.8 & 1.02 & 10.08\\
  -1.05 & 45.95 & 0.72 & 10.01\\
  -1.92 & 45.88 & 0.79 & 5.04\\
\hline
\end{tabular}
\end{table}

\subsection{Construction of the SDSS background} \label{background}

Since the overdensities found in previous studies are clearly visible along the stream, we next investigate whether it is also possible to detect similar features in the absence of a stream, thus if their presence may be an artefact of the survey and detection method. To be certain that overdensities detected in the background will not interfere with our analysis, we constructed three different realisations of it.

To create these models of the background in the SDSS, we remove all of the data within $1 \degr$ around the stream; this is appropriate since the stream is very thin \citep[$\sim 0.2 \degr$ , see ][]{odenkirchen_2003}. Then, we fill this area by duplicating data from an adjacent region between respectively $1 \degr$, $1.5 \degr$ and $2 \degr$ to the North for the leading arm, and $1 \degr$, $1.5 \degr$ and $2 \degr$  to the South for the tailing arm, and deredden the stars with the extinction value of their new position. This very simple method allows us to re-use the pipeline developed in Section \ref{def_method}. This of course assumes that the statistics of the background under the stream are similar to those in the adjacent field.

\subsection{Model of a smooth stream} \label{smooth _stream}

To understand better the limitation of the detection of overdensities, we created a smooth model of the stream that we added to our background models, previously described. 

First, we determine the number of stars detected by the matched-filter method along the stream in the SDSS. To this end, we select the stars in an area with thickness of 0.4 deg along the stream, assuming that the position of the center of the Pal 5 stream can be determined by the following functions in the standard coordinates $(\xi, \eta)$ \citep{ibata_2016}, where $(\xi, \eta) = (0,0)$ corresponds to the center of the globular cluster:

  \begin{equation}
 \left.
  \begin{array}{ l }
\eta_{trailing}(\xi)= 0.211 + 0.768 \xi -0.0305 \xi^2 + 0.000845 \xi^3\\

\eta_{leading}(\xi)= -0.199 + 0.919 \xi +0.0226 \xi^2 + 0.0123 \xi^3 \, .
 \end{array}
  \right.
  \label{position_stream}
\end{equation}

We masked out a region of 1 deg around the globular cluster, since we are only interested in the properties of the stream.
We found 1570 stream stars within a full-width of 0.4 degrees around the above cubic polynomial model in the matched-filter map, which we corrected to 1805 stars to make a smooth stream and fill the masked region around the globular cluster, assuming that the density surface is constant along the stream. To obtain a perfectly smooth model, these stars were distributed randomly along the stream, with a full thickness of 0.4 deg around the center of the stream.
As in Section \ref{def_method}, we used the data from the center of the globular cluster to determine the probability function of the luminosity in the \rm{g}, \rm{r} and \rm{i}-bands. Thus we assigned random magnitudes to the stars of our model, following these functions, we applied the completeness function determined in Section \ref{CFHT}, before dereddening them.

\subsection{Nbody simulation of the stream} \label{Nbody}

To create dynamically plausible stream models, we also undertook N-body simulations of the disruption of the globular cluster Pal 5, using the GyrfalcON integrator \citep{dehnen_2000} from the NEMO package \citep{teuben_1995}.

In our simulation, the distribution of the baryonic matter of the Milky Way is modelled with the bulge, the thin and thick disks and the ISM component defined in the 1$^{st}$ model of \cite{dehnen_1998}. However, like \cite{kupper_2015}, we prefer to model the dark matter halo with a Navarro-Frenk-White distribution \citep[NFW;][]{navarro_1997} with an oblateness along the axis perpendicular to the Galactic disk of $q_z = 0.94$, a virial mass of $M_{200}=1.60 \times 10^{12} M_\odot$, a scale length of $r_s=36.5$ kpc and a concentration of $c=5.95$.

Our model of the progenitor of Pal 5 follows a King model with a mass $M_{gc}=2 \times 10^4$ M$_\odot$, a core radius of $r_{0,gc}=50$~pc and a ratio between the central potential and the velocity dispersion of $W_{gc}=2.5$. We use the current parameters of the cluster, listed in the Table \ref{param_pal5}, as the required final state of the progenitor in the simulation. The current tangential velocity of the globular cluster was determined by running 200 N-body simulations to fit the position of the centre of the simulated stream to Equation \ref{position_stream} and to fit the radial velocity along the stream to the observed radial velocities of \cite{odenkirchen_2009}. We find a tangential velocity of $V_{tan}=335$ km s$^{-1}$, consistent with the recent measurement of \cite{fritz_2015} obtained with the Hubble Space Telescope.

 \begin{table}
 \centering
  \caption{Properties of the globular cluster Pal 5. The sources are : $1 =$ \citet{dicriscienzo_2006}, $2 =$ \citet{ibata_2016}, $3 =$ \citet{odenkirchen_2002}, $4 =$ \citet{smith_2002}.}
  \label{param_pal5}
  \begin{tabular}{@{}ccc@{}}
  \hline
   Parameter & Value & Source  \\
    \hline
   RA & $15^h16^m5.3^s$ & 1 \\
   Dec & $-00\degr 06'41.0"$ & 1 \\
   Distance & $23.5$ kpc & 2 \\
   V$_{rad}$ & $-58.7 \pm 0.2$ km.s$^{-1}$ & 3\\
   $\mu_\alpha$ & -2.13 mas.yr$^{-1}$&  \\
   $\mu_\delta$ & -2.13 mas.yr$^{-1}$&  \\
   $[$Fe/H$]$ & $-1.3$ & 4 \\
   Age & $11.5$ Gyr & \\
\hline
\end{tabular}
\end{table}

We adopted a smoothing scale length in GyrfalcON of $0.3$ pc, and chose to simulate the globular cluster with $2 \times 10^5$ equal-mass particles. However, by changing the number of particles and the smoothing length, with $2 \times 10^4$ N-body particles and a smoothing length of 3 pc, we checked that the morphology of the stream does not depend significantly on these choices. 

We transformed these N-body particles into stellar particles with a range of stellar mass and observable properties, assuming a Salpeter mass function, and a magnitude in the \rm{g}, \rm{r} and \rm{i} bands drawn from an isochrone of age $11.5$ Gyr and a metallicity of $[$Fe/H$] = -1.3$ from the Dartmouth stellar tracks \citep{dotter_2008}. However, the number of stellar particles that are bright enough to be detected in the SDSS is smaller than observed, thus we limit our mass distribution at $0.5$ M$_\odot$, which corresponds to an absolute magnitude of $8.5$ in the \rm{g}-band. This effect is not a consequence of the Salpeter mass function, since \cite{kupper_2015} also found themselves forced to truncate their adopted \cite{kroupa_2001} mass function at  $0.5$ M$_\odot$.

Finally, to have the same observational biases as the true stream in the SDSS, we add the extinction to the stars and apply the completeness determined in Section \ref{CFHT}. However, we note that here we are assuming that we know the extinction perfectly, as we use precisely the same values that were added to account for the reddening of the stars, therefore the only effect of the extinction on the simulations is to correct for the completeness.

\section{Results} \label{result}

\begin{figure*}
  \includegraphics[angle=0, viewport= 10 30 565 273, clip, width=15cm]{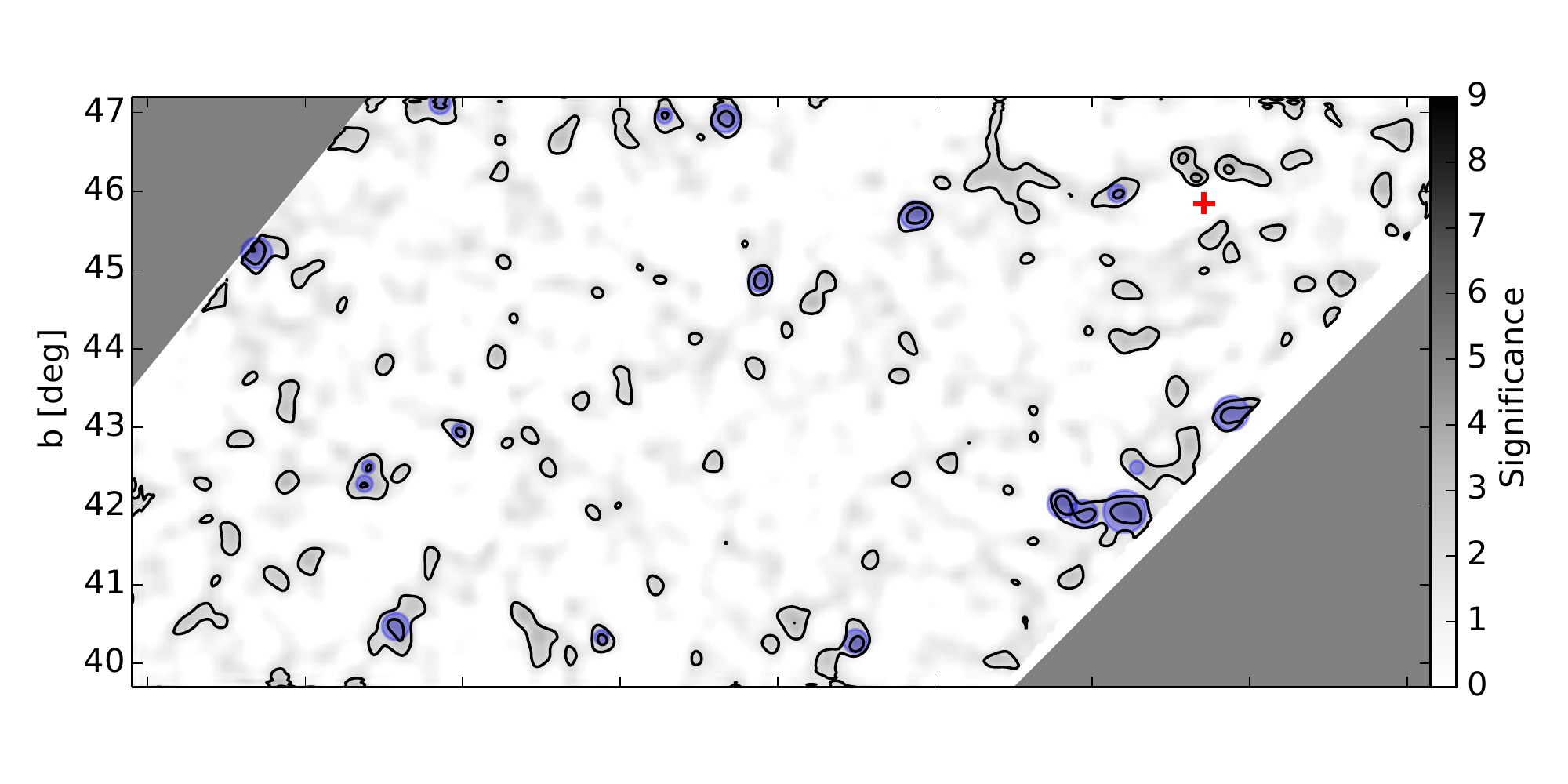}
  \includegraphics[angle=0, viewport= 10 30 565 263, clip, width=15cm]{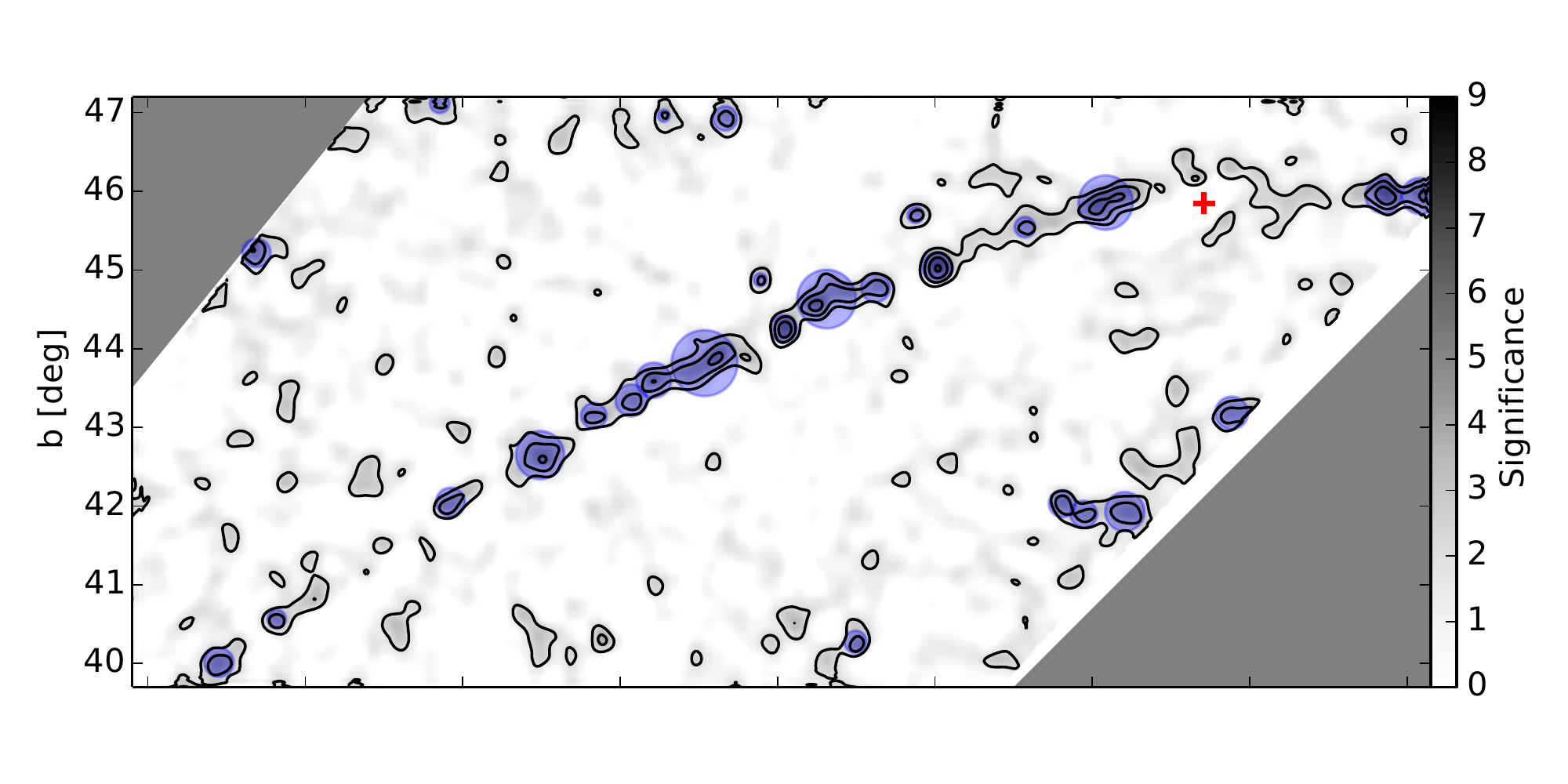}
  \includegraphics[angle=0, viewport= 8 15 605 287, clip, width=15cm]{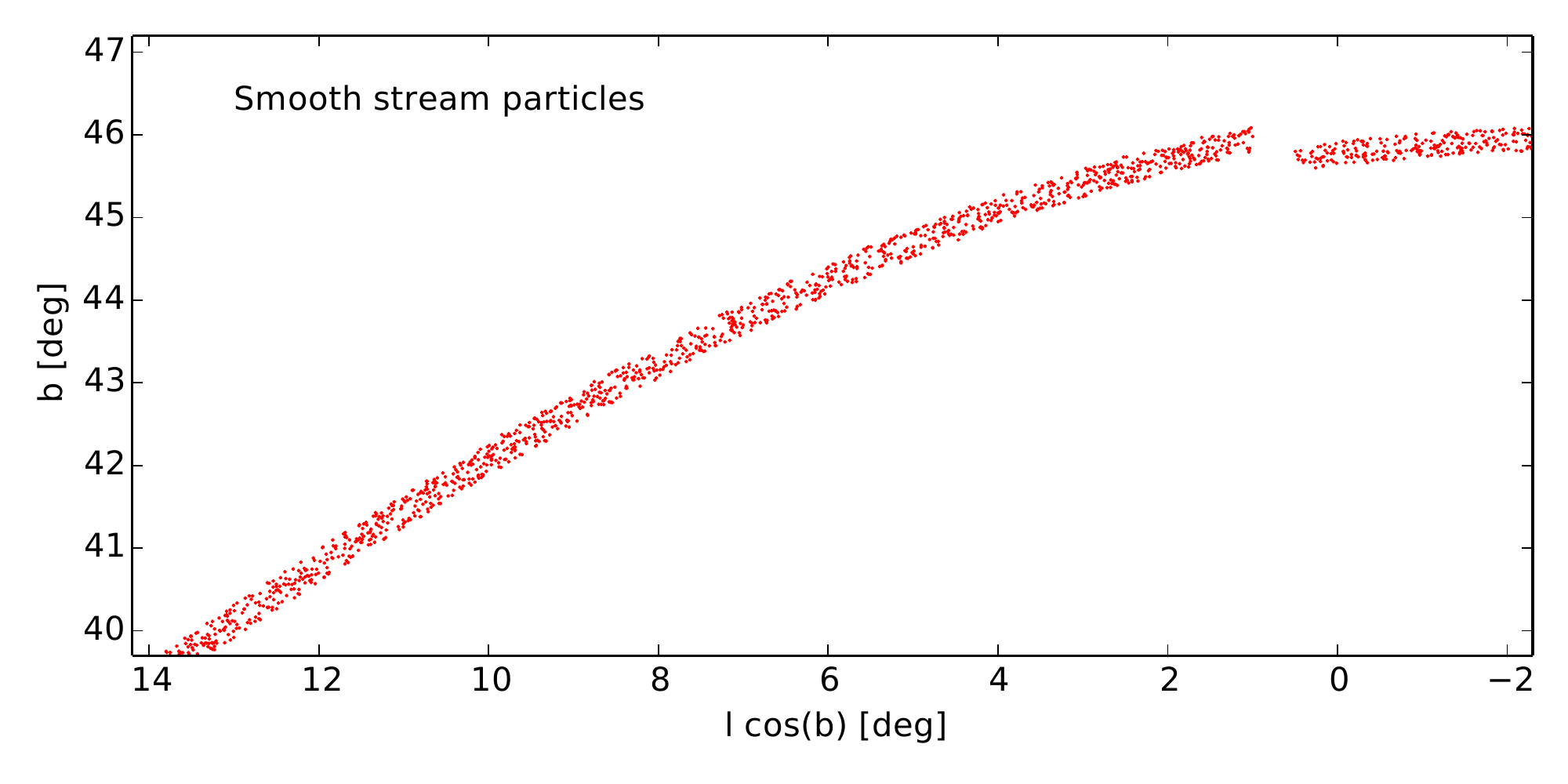}

  \caption{As Figure \ref{sig_SDSS}, but with the significance map of the first model of the ``SDSS background'' shown on the top panel, while in the middle panel we show the significance map of the 1805 stars from the smooth stream added to that model of the ``SDSS background''. In the lower panel we show the position of the stellar particles of the smooth stream. The gaps between $0.5\degr < l\, \cos(b) < 1.0\degr$ correspond to the region at the center of the globular cluster that we do not use in this study. }
\label{sig_lisse}
\end{figure*}

\subsection{Analysis of the observed substructures}
As explained in Section \ref{def_method}, we applied the method described above to find the overdensities in the region around the stream of Pal5 in the SDSS. Thus, the topography of the significance is shown in Figure \ref{sig_SDSS}, where the blue circles represent the position of the overdensities and their radius is equal to the FWHM of the overdensities given by SExtractor. Their mean significance over the whole field is of $5.80$, while that of the overdensities along the stream is higher with $S_{mean,obs}=6.70$.

It is obvious that in the region between $6\degr < l\, \cos(b) < 10\degr$, there is an absence of overdensities. It is in this region that the extinction along the stream is higher (Figure \ref{extinction}), with a mean extinction in the \rm{g}-band of $A_g \sim 0.22$. 

However, in the deeper observations from the CFHT \citep{ibata_2016}, and in N-body simulations \citep{dehnen_2004, kupper_2010}, the density along the Pal 5 stream decreases approximately smoothly with distance and there is an absence of obvious density variations, excepting those probably induced by the epicyclic motion of stars along the stream. Thus, it is natural to wonder if the overdensities detected along the Pal 5 stream in the SDSS are physical or if they are due to the inhomogeneity of the SDSS survey and the matched filter method that was used to detect the substructures.

\subsection{Substructures in the smooth stream}
To answer this question, we search for overdensities in the artificial smooth stream, represented in the middle panel of Figure~\ref{sig_lisse} (Section \ref{smooth _stream}). Since the input stream is perfectly smooth within Poisson uncertainties, one would expect to have approximately constant significance along it, which should be higher than the significance of the background, and an absence of overdensities.

From the lower panel of the Figure \ref{sig_lisse}, where the overdensities are detected from the smooth stream added to the that ``background SDSS'', it is obvious that the stream, seen in the same way as the observations, is fragmented and seems to have great variations in density along it, contrary to expectation given that the input model is a smooth stream.

As the smooth stream is added to the real SDSS background, which can introduce substructures in the stream that are already present in the background, we analyse next the overdensities detected in the background. The top panel of Figure \ref{sig_lisse} presents the spatial distribution of the overdensities of our first model of the background component, which have a mean significance of $S_{mean,bg}=4.08$, well under those of the observation ($S_{mean,obs}$). Moreover, the overdensities are spread approximately homogeneously along the stream and their positions are different than those detected along the smooth stream, so their impact on it will be minimal, validating our method of constructing this background. Thus even if the background contains the ``seeds'' of the overdensities detected along the smooth stream, they are not the principal contributor to them and we can suppose that a similar phenomenon is also present with the overdensities detected in the real SDSS observations.

\begin{figure}
  \includegraphics[scale=0.44]{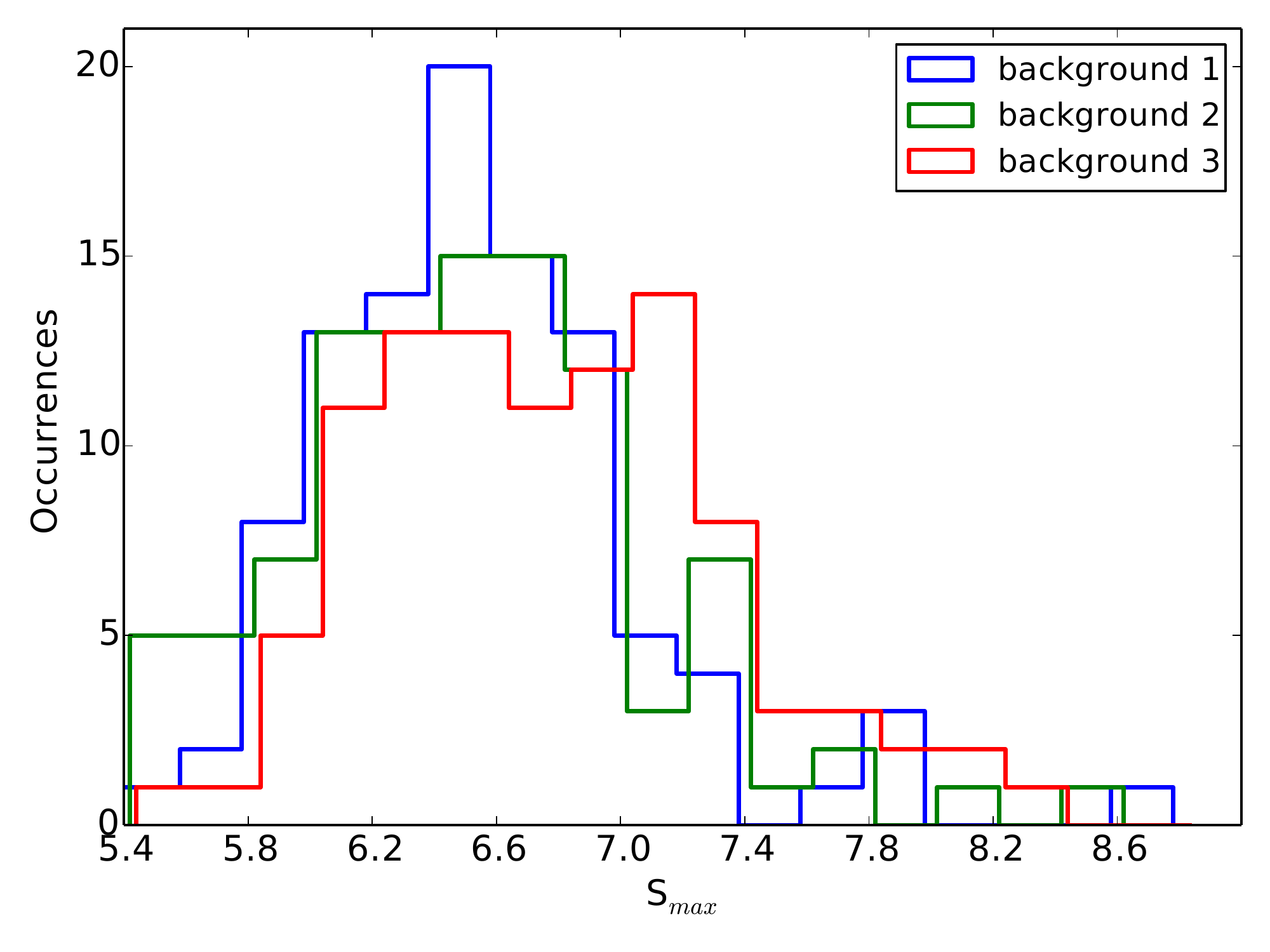}
  \caption{ Histogram of the maximum $S$ value measured in 100 random realizations of our smooth stream model with the three models of the background.} 
\label{Smax}
\end{figure}

To make sure that the presence of these false overdensities is not an unlucky consequence of the particular random number seed used in generating Figure \ref{sig_lisse}, we constructed a sample of 100 realisations of this smooth model for the three models of the background  previously mentioned (see Section \ref{background}), always with the same number of stars in the stream, and we measured the maximum significance of the peaks in these models. Figure \ref{Smax} shows the histogram of the maximum significance of the detected peaks, and shows overdensities along the smooth stream reaching maximum values of $S=8.74$, 8.61 and  8.25 for background models 1, 2 and 3, respectively. Thus we conclude that any overdensities in the real observations with $S$ \simlt $7.8$ are suspicious and may not be real, since similar peaks arise by chance in 5 out of 100 random realisations of a smooth stream. With this criterion there are five overdensities detected in the SDSS stream, listed in Table \ref{table_SDSS}, that could be considered significant.

\subsection{Substructures in the simulated stream}

\begin{figure*}
  \includegraphics[angle=0, viewport= 8 45 606 277, clip, width=15cm]{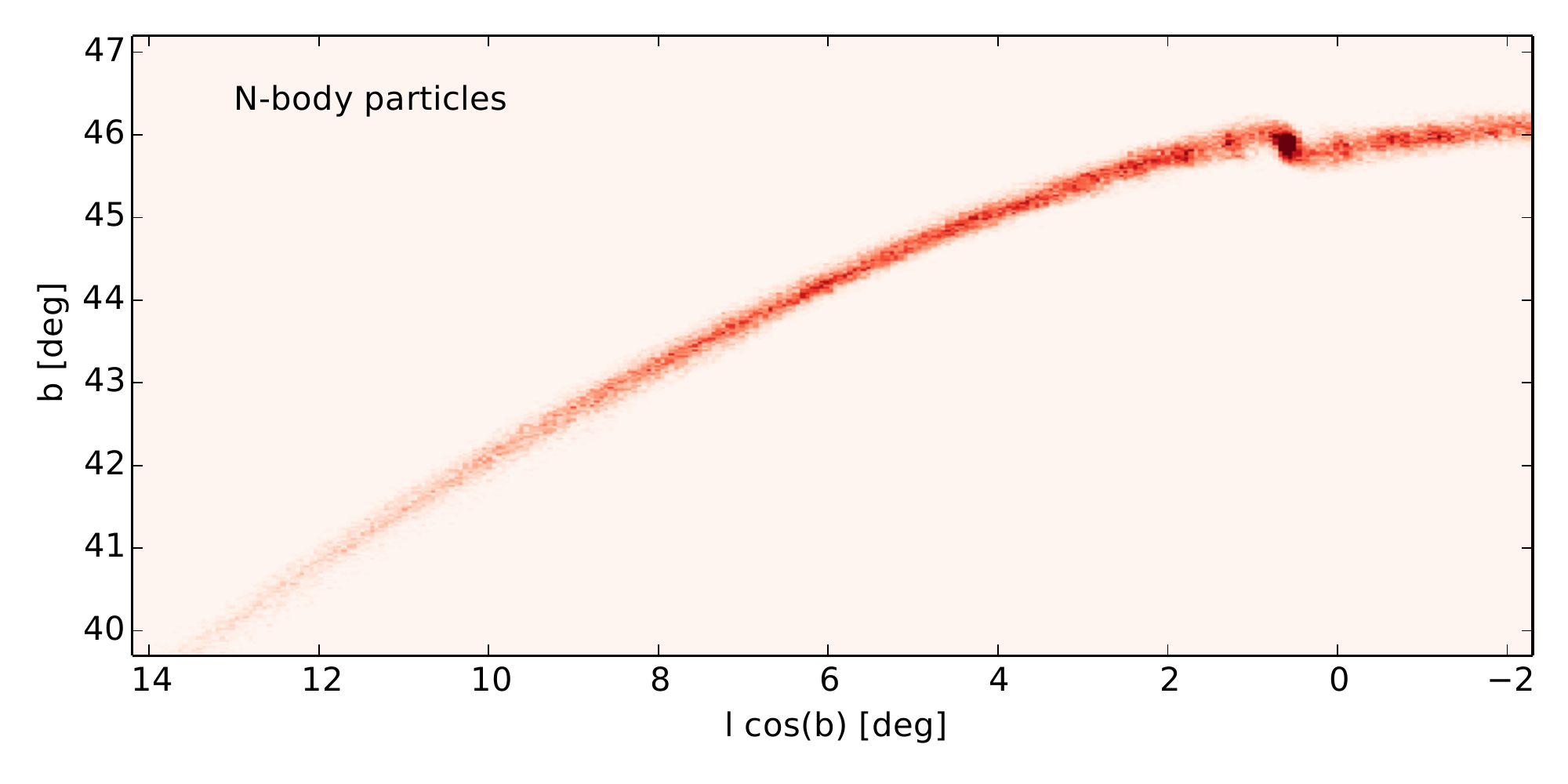}
  \includegraphics[angle=0, viewport= 8 45 606 287, clip, width=15cm]{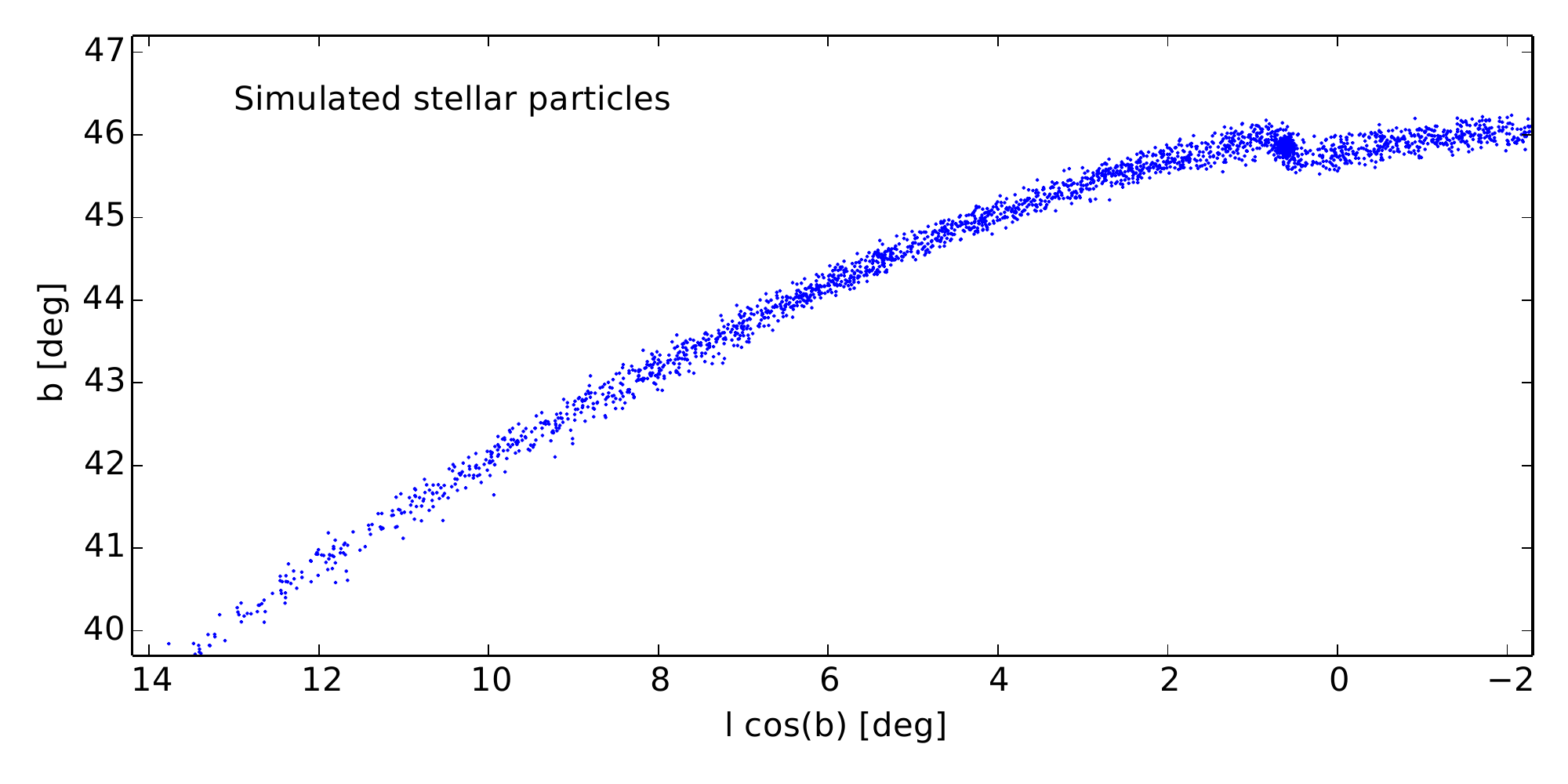}
  \includegraphics[angle=0, viewport= 10 30 565 273, clip, width=15cm]{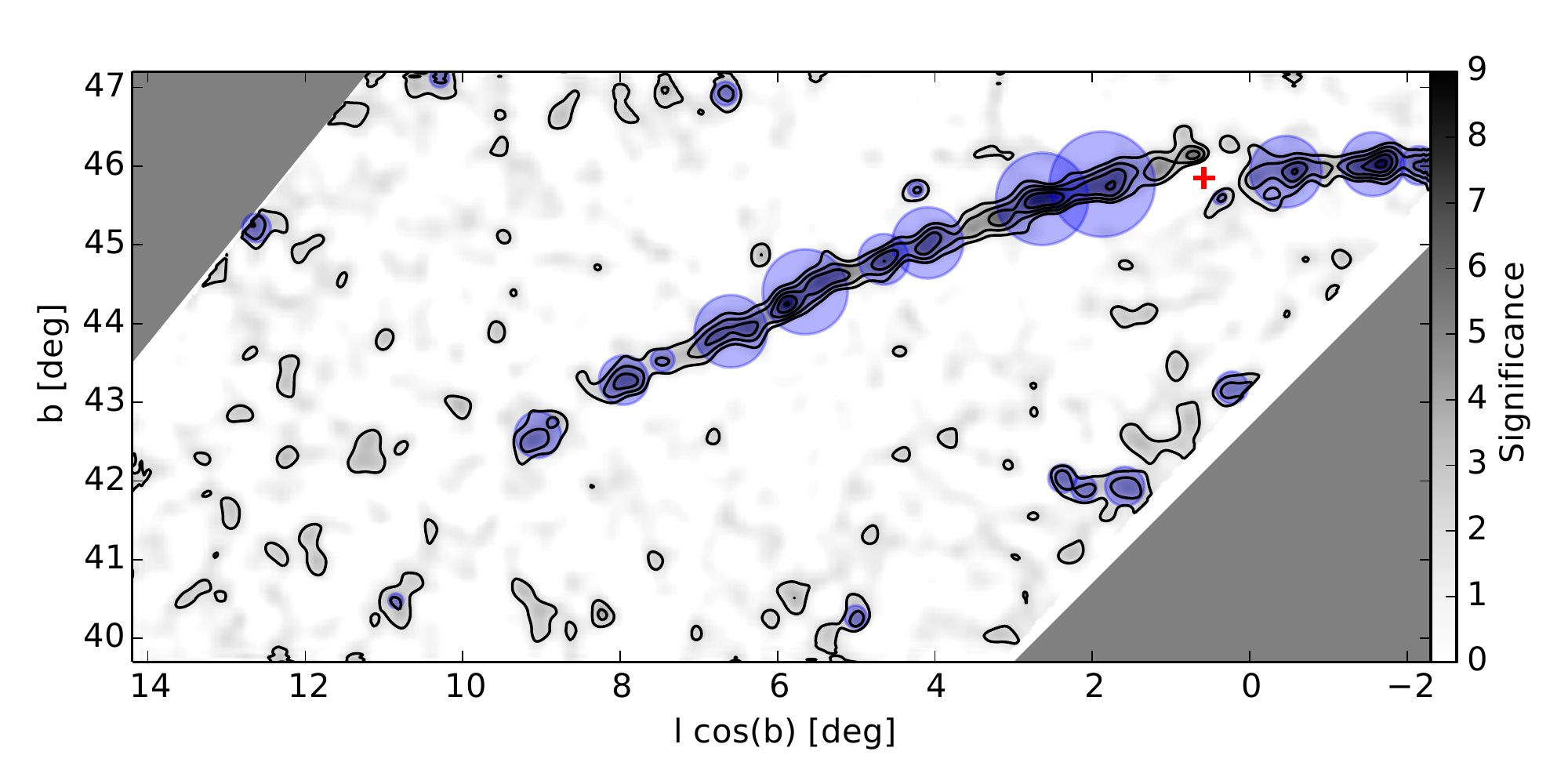}
  \caption{ Position of the $2 \times 10^5$ N-body particles (in red on the top panel) and of the $6000$ simulated stellar particles that respect the SDSS star-counts constraint (in blue on the middle panel). The position of the overdensities detected along the simulated stream is represented on the bottom panel. }
\label{sig_simu}
\end{figure*}

We also ran N-body simulations of the disruption of the globular cluster Pal 5 in order to reproduce the observed stream (see Section \ref{Nbody}). The best-fit model stream, once the extinction and the completeness are accounted for, is composed of $\sim 6000$ stellar particles, of which approximately 70\% stay bound to the globular cluster. However, using the same method as in Section \ref{smooth _stream} to extract the number of stars along the stream from the matched-filter map, we find $1580$ stellar particles, which is in agreement with the star-count constraint from the SDSS. Figure \ref{sig_simu} shows, on the top panel, the initial $2 \times 10^5$ particles used to follow the dynamics of the stream, while on the middle panel we present the $\sim 6000$ star-particles that compose the stream once the SDSS observational biases are added. 

\begin{figure}
  \includegraphics[scale=0.45]{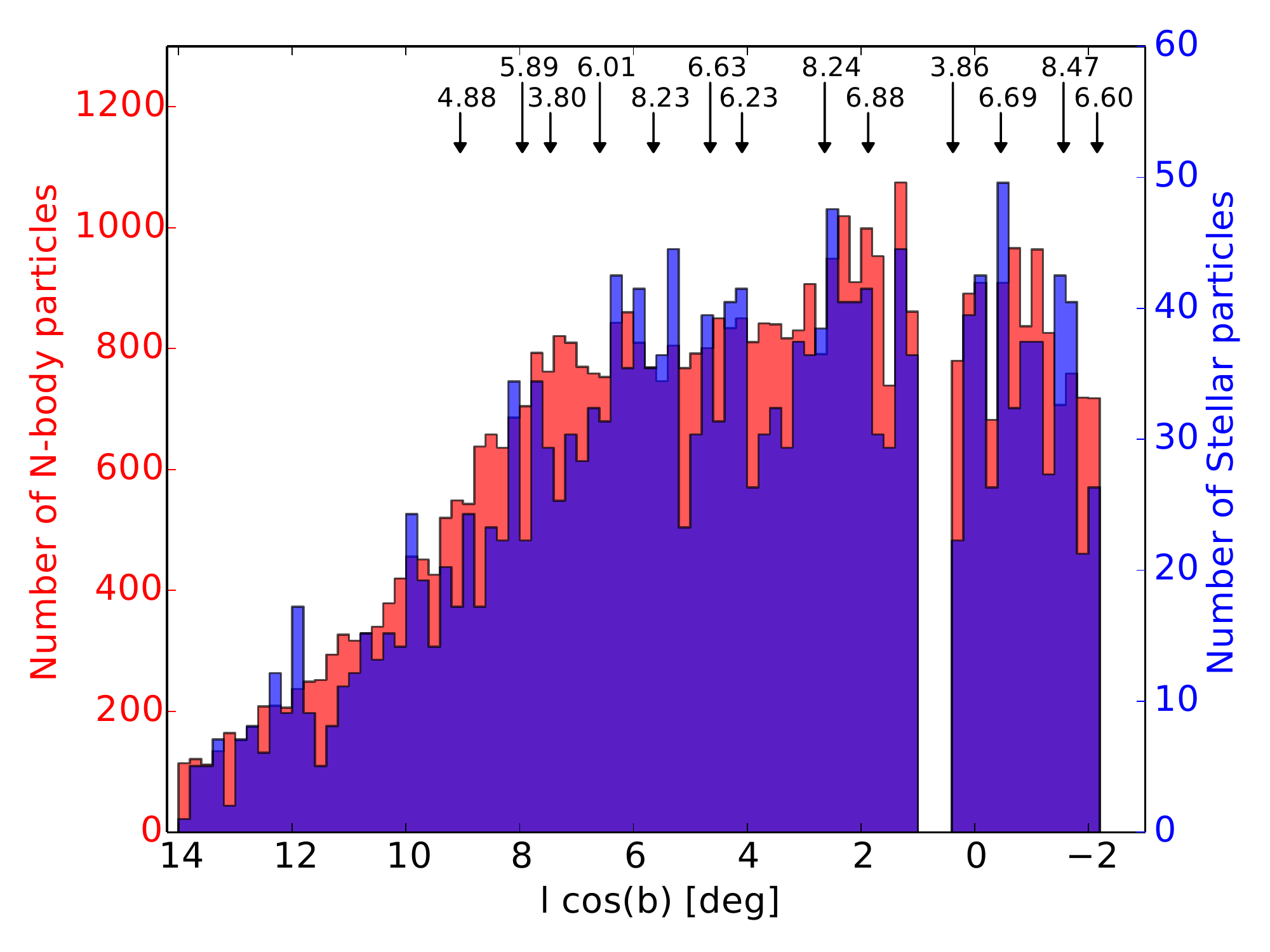}
  \caption{ Density of N-body particles (in red) and stellar particles (in blue) along the simulated stream. The vertical lines show the position of the overdensities detected with the method previously mentioned, and the corresponding ``significance'' statistic of the peak is indicated.}
\label{density_simu}
\end{figure}

The density variation that can be seen along the N-body stream in the top panel is a natural consequence of the epicyclic motion performed by a continuous flux of stars that escape from the satellite \citep{kupper_2012}. However, this phenomenon is less clearly visible in the stream formed out of star particles, simply due to the much smaller number of SDSS stars actually available, and as we show in Figure \ref{density_simu}, the positions of the peaks in the star particle distribution (blue histogram and arrows) do not unambiguously reveal the positions of the true N-body peaks (red histogram): while some peaks do match up, especially those closest to the cluster, many high significance peaks can be seen to be artefacts of the method, with as high a significance as for the true peaks.

The bottom panel of Figure \ref{sig_simu}, similar to Figure \ref{sig_SDSS}, shows the location of the overdensities detected along the stream using the method adopted in this study. The mean value of the significance along the stream is very close to S$_{mean,obs}$ with S$_{mean,simu} = 6.67$.  

It is interesting to note that the simulated N-body stream is detected over the region $6\degr < l\, \cos(b) < 10\degr$, where the observed stream shows little signal, and extinction is higher than average (especially for $8.5\degr < l\, \cos(b) < 10\degr$). This suggests that the relatively small variations in extinction in the Palomar 5 field do not affect substantially the detectability of the stream.

\section{Discussion and conclusions} \label{conclusion}

Various processes have been proposed to explain the great variation in density seen in the SDSS along the tidal tails of the globular cluster Palomar 5. The options include violent tidal shocks as the cluster passes through the Milky Way disk, or close to dense star-forming regions, the epicyclic motion of stars emanating from the cluster, or encounters with dark matter subhalos. However, the absence of similar substructures in the recent deeper CFHT observations presented by \cite{ibata_2016} led us to question their reality, and raised the possibility that they are artefacts generated by the particular photometric depth and inhomogeneity of the SDSS linked to the analysis techniques used to detect these very low contrast features.

To answer this question we have reproduced a similar method as \cite{kupper_2015} to extract the position of overdensities along the stellar stream of Pal 5. Firstly, we applied this method to the SDSS DR9 and found many overdensities that do not overlap. The apparently high significance of these substructures seems consistent with previous claims that there are strong variations in the density of stars along the stream.

We subsequently applied the same algorithm to a sample of 100 random realisations of a smooth stream and found in every realisation apparently significant overdensities that were not due to the modelled background. This study has also shown that overdensities with a significance (as defined above) below $\sim 7$ are highly questionable to be a signature of real density variations.

We have also undertaken N-body simulations, to fit as closely as possible the disruption of Pal 5. In our best-fit model, the variations of the density of stars generated by the epicyclic motion is clearly visible in the N-body particles. However, this effect disappears once these N-body particles are transformed into stellar particles and the SDSS photometric selection is applied, due to the small number of particles that remain.

Finally, we can conclude that the variations in density seen along the stream of Palomar 5 are largely due to the effect of the small number statistics in the SDSS.  This problem is compounded by the properties of the matched-filter technique which weights stars according to their CMD location, thus boosting certain stars and thereby biassing usual significance metrics. This is especially problematic for a survey like the SDSS where the faint stars with large photometric uncertainties are not easily differentiable from the background. However, it is surprising to find little or no correlation between stream density and extinction, though this may be due simply to the relatively small and limited range of reddening  along the stream.

In future work it will be interesting to ascertain whether a similar effect could explain the gaps seen along other stellar streams, such as GD-1 \citep{carlberg_2013}.

\medskip
The authors would like to thank Andreas K\"upper and Eduardo Balbinot for very helpful discussions regarding the implementation of their substructure detection method.

Funding for SDSS-III has been provided by the Alfred P. Sloan Foundation, the Participating Institutions, the National Science Foundation, and the U.S. Department of Energy Office of Science. The SDSS-III web site is http://sdss3.org.

Based on observations obtained with MegaPrime/MegaCam, a joint project of CFHT and CEA/DAPNIA, at the Canada-France-Hawaii Telescope (CFHT) which is operated by the National Research Council (NRC) of Canada, the Institute National des Sciences de l'Univers of the Centre National de la Recherche Scientifique of France, and the University of Hawaii.

\bibliography{./biblio}

\begin{thebibliography}{}

\bibitem[\protect\citeauthoryear{Ahn, Alexandroff, {Allende Prieto}, Anderson,
  Anderton, Andrews, Aubourg, Bailey \& {et al,}}{Ahn et~al.}{2012}]{ahn_2012}
Ahn C.~P.,  Alexandroff R.,  {Allende Prieto} C.,  Anderson S.~F.,  Anderton
  T.,  Andrews B.~H.,  Aubourg {\'E}.,  Bailey S.,    {et al,} 2012, The
  Astrophysical Journal Supplement Series, 203, 21

\bibitem[\protect\citeauthoryear{Balbinot, Santiago, {da Costa}, Makler \&
  Maia}{Balbinot et~al.}{2011}]{balbinot_2011}
Balbinot E.,  Santiago B.~X.,  {da Costa} L.~N.,  Makler M.,    Maia M. A.~G.,
  2011, Monthly Notices of the Royal Astronomical Society, 416, 393

\bibitem[\protect\citeauthoryear{Belokurov, Zucker, Evans, Gilmore, Vidrih,
  Bramich, Newberg, Wyse, Irwin, Fellhauer, Hewett, Walton, Wilkinson, Cole,
  Yanny, Rockosi, Beers, Bell, Brinkmann, Ivezi{\'c} \& Lupton}{Belokurov
  et~al.}{2006}]{belokurov_2006}
Belokurov V.,  Zucker D.~B.,  Evans N.~W.,  Gilmore G.,  Vidrih S.,  Bramich
  D.~M.,  Newberg H.~J.,  Wyse R. F.~G.,  Irwin M.~J.,  Fellhauer M.,  Hewett
  P.~C.,  Walton N.~A.,  Wilkinson M.~I.,  Cole N.,  Yanny B.,  Rockosi C.~M.,
  Beers T.~C.,  Bell E.~F.,  Brinkmann J.,  Ivezi{\'c} {\v Z}.,    Lupton R.,
  2006, The Astrophysical Journal Letters, 642, L137

\bibitem[\protect\citeauthoryear{Bertin \& Arnouts}{Bertin \&
  Arnouts}{1996}]{bertin_1996}
Bertin E.,  Arnouts S.,  1996, Astronomy and Astrophysics Supplement Series,
  117, 393

\bibitem[\protect\citeauthoryear{Carlberg}{Carlberg}{2012}]{carlberg_2012a}
Carlberg R.~G.,  2012, The Astrophysical Journal, 748, 20

\bibitem[\protect\citeauthoryear{Carlberg \& Grillmair}{Carlberg \&
  Grillmair}{2013}]{carlberg_2013}
Carlberg R.~G.,  Grillmair C.~J.,  2013, The Astrophysical Journal, 768, 171

\bibitem[\protect\citeauthoryear{Carlberg, Grillmair \& Hetherington}{Carlberg
  et~al.}{2012}]{carlberg_2012}
Carlberg R.~G.,  Grillmair C.~J.,    Hetherington N.,  2012, The Astrophysical
  Journal, 760, 75

\bibitem[\protect\citeauthoryear{Dehnen}{Dehnen}{2000}]{dehnen_2000}
Dehnen W.,  2000, The Astrophysical Journal Letters, 536, L39

\bibitem[\protect\citeauthoryear{Dehnen \& Binney}{Dehnen \&
  Binney}{1998}]{dehnen_1998}
Dehnen W.,  Binney J.,  1998, Monthly Notices of the Royal Astronomical
  Society, 294, 429

\bibitem[\protect\citeauthoryear{Dehnen, Odenkirchen, Grebel \& Rix}{Dehnen
  et~al.}{2004}]{dehnen_2004}
Dehnen W.,  Odenkirchen M.,  Grebel E.~K.,    Rix H.-W.,  2004, The
  Astronomical Journal, 127, 2753

\bibitem[\protect\citeauthoryear{{Di Criscienzo}, Caputo, Marconi \&
  Musella}{{Di Criscienzo} et~al.}{2006}]{dicriscienzo_2006}
{Di Criscienzo} M.,  Caputo F.,  Marconi M.,    Musella I.,  2006, Monthly
  Notices of the Royal Astronomical Society, 365, 1357

\bibitem[\protect\citeauthoryear{Dotter, Chaboyer, Jevremovi{\'c}, Kostov,
  Baron \& Ferguson}{Dotter et~al.}{2008}]{dotter_2008}
Dotter A.,  Chaboyer B.,  Jevremovi{\'c} D.,  Kostov V.,  Baron E.,    Ferguson
  J.~W.,  2008, The Astrophysical Journal Supplement Series, 178, 89

\bibitem[\protect\citeauthoryear{Fritz \& Kallivayalil}{Fritz \&
  Kallivayalil}{2015}]{fritz_2015}
Fritz T.~K.,  Kallivayalil N.,  2015, The Astrophysical Journal, 811, 123

\bibitem[\protect\citeauthoryear{Grillmair}{Grillmair}{2006}]{grillmair_2006b}
Grillmair C.~J.,  2006, The Astrophysical Journal Letters, 645, L37

\bibitem[\protect\citeauthoryear{Grillmair}{Grillmair}{2009}]{grillmair_2009}
Grillmair C.~J.,  2009, The Astrophysical Journal, 693, 1118

\bibitem[\protect\citeauthoryear{Grillmair \& Dionatos}{Grillmair \&
  Dionatos}{2006a}]{grillmair_2006}
Grillmair C.~J.,  Dionatos O.,  2006a, The Astrophysical Journal Letters, 641,
  L37

\bibitem[\protect\citeauthoryear{Grillmair \& Dionatos}{Grillmair \&
  Dionatos}{2006b}]{grillmair_2006a}
Grillmair C.~J.,  Dionatos O.,  2006b, The Astrophysical Journal Letters, 643,
  L17

\bibitem[\protect\citeauthoryear{Ibata, Lewis, Irwin, Totten \& Quinn}{Ibata
  et~al.}{2001}]{ibata_2001}
Ibata R.,  Lewis G.~F.,  Irwin M.,  Totten E.,    Quinn T.,  2001, The
  Astrophysical Journal, 551, 294

\bibitem[\protect\citeauthoryear{Ibata, Ibata, Lewis, Martin, Conn, Elahi,
  Arias \& Fernando}{Ibata et~al.}{2014}]{ibata_2014b}
Ibata R.~A.,  Ibata N.~G.,  Lewis G.~F.,  Martin N.~F.,  Conn A.,  Elahi P.,
  Arias V.,    Fernando N.,  2014, The Astrophysical Journal Letters, 784, L6

\bibitem[\protect\citeauthoryear{Ibata, Lewis, Irwin \& Quinn}{Ibata
  et~al.}{2002}]{ibata_2002}
Ibata R.~A.,  Lewis G.~F.,  Irwin M.~J.,    Quinn T.,  2002, Monthly Notices of
  the Royal Astronomical Society, 332, 915

\bibitem[\protect\citeauthoryear{Ibata, Lewis \& Martin}{Ibata
  et~al.}{2016}]{ibata_2016}
Ibata R.~A.,  Lewis G.~F.,    Martin N.~F.,  2016, The Astrophysical Journal,
  819, 1

\bibitem[\protect\citeauthoryear{Johnston, Spergel \& Haydn}{Johnston
  et~al.}{2002}]{johnston_2002}
Johnston K.~V.,  Spergel D.~N.,    Haydn C.,  2002, The Astrophysical Journal,
  570, 656

\bibitem[\protect\citeauthoryear{Koposov, Glushkova \& Zolotukhin}{Koposov
  et~al.}{2008}]{Koposov_2008}
Koposov S.~E.,  Glushkova E.~V.,    Zolotukhin I.~Y.,  2008, Astronomy and
  Astrophysics, 486, 771

\bibitem[\protect\citeauthoryear{Kroupa}{Kroupa}{2001}]{kroupa_2001}
Kroupa P.,  2001, Monthly Notices of the Royal Astronomical Society, 322, 231

\bibitem[\protect\citeauthoryear{K{\"u}pper, Balbinot, Bonaca, Johnston, Hogg,
  Kroupa \& Santiago}{K{\"u}pper et~al.}{2015}]{kupper_2015}
K{\"u}pper A. H.~W.,  Balbinot E.,  Bonaca A.,  Johnston K.~V.,  Hogg D.~W.,
  Kroupa P.,    Santiago B.~X.,  2015, The Astrophysical Journal, 803, 80

\bibitem[\protect\citeauthoryear{K{\"u}pper, Kroupa, Baumgardt \&
  Heggie}{K{\"u}pper et~al.}{2010}]{kupper_2010}
K{\"u}pper A. H.~W.,  Kroupa P.,  Baumgardt H.,    Heggie D.~C.,  2010, Monthly
  Notices of the Royal Astronomical Society, 401, 105

\bibitem[\protect\citeauthoryear{K{\"u}pper, Lane \& Heggie}{K{\"u}pper
  et~al.}{2012}]{kupper_2012}
K{\"u}pper A. H.~W.,  Lane R.~R.,    Heggie D.~C.,  2012, Monthly Notices of
  the Royal Astronomical Society, 420, 2700

\bibitem[\protect\citeauthoryear{K{\"u}pper, MacLeod \& Heggie}{K{\"u}pper
  et~al.}{2008}]{kupper_2008}
K{\"u}pper A. H.~W.,  MacLeod A.,    Heggie D.~C.,  2008, Monthly Notices of
  the Royal Astronomical Society, 387, 1248

\bibitem[\protect\citeauthoryear{Mart{\'\i}nez-Delgado, Gabany, Crawford,
  Zibetti, Majewski, Rix, Fliri, Carballo-Bello, Bardalez-Gagliuffi,
  Pe{\~n}arrubia, Chonis, Madore, Trujillo, Schirmer \&
  McDavid}{Mart{\'\i}nez-Delgado et~al.}{2010}]{martinez-delgado_2010}
Mart{\'\i}nez-Delgado D.,  Gabany R.~J.,  Crawford K.,  Zibetti S.,  Majewski
  S.~R.,  Rix H.-W.,  Fliri J.,  Carballo-Bello J.~A.,  Bardalez-Gagliuffi
  D.~C.,  Pe{\~n}arrubia J.,  Chonis T.~S.,  Madore B.,  Trujillo I.,  Schirmer
  M.,    McDavid D.~A.,  2010, The Astronomical Journal, 140, 962

\bibitem[\protect\citeauthoryear{Monari, Famaey \& Siebert}{Monari
  et~al.}{2016}]{monari_2016}
Monari G.,  Famaey B.,    Siebert A.,  2016, Monthly Notices of the Royal
  Astronomical Society, 457, 2569

\bibitem[\protect\citeauthoryear{Navarro, Frenk \& White}{Navarro
  et~al.}{1997}]{navarro_1997}
Navarro J.~F.,  Frenk C.~S.,    White S. D.~M.,  1997, The Astrophysical
  Journal, 490, 493

\bibitem[\protect\citeauthoryear{Ngan, Bozek, Carlberg, Wyse, Szalay \&
  Madau}{Ngan et~al.}{2014}]{ngan_2014}
Ngan W.,  Bozek B.,  Carlberg R.~G.,  Wyse R. F.~G.,  Szalay A.~S.,    Madau
  P.,  2014, ArXiv e-prints, 1411, 3760

\bibitem[\protect\citeauthoryear{Odenkirchen, Grebel, Dehnen, Rix \&
  Cudworth}{Odenkirchen et~al.}{2002}]{odenkirchen_2002}
Odenkirchen M.,  Grebel E.~K.,  Dehnen W.,  Rix H.-W.,    Cudworth K.~M.,
  2002, The Astronomical Journal, 124, 1497

\bibitem[\protect\citeauthoryear{Odenkirchen, Grebel, Dehnen, Rix, Yanny,
  Newberg, Rockosi, Mart{\'\i}nez-Delgado, Brinkmann \& Pier}{Odenkirchen
  et~al.}{2003}]{odenkirchen_2003}
Odenkirchen M.,  Grebel E.~K.,  Dehnen W.,  Rix H.-W.,  Yanny B.,  Newberg
  H.~J.,  Rockosi C.~M.,  Mart{\'\i}nez-Delgado D.,  Brinkmann J.,    Pier
  J.~R.,  2003, The Astronomical Journal, 126, 2385

\bibitem[\protect\citeauthoryear{Odenkirchen, Grebel, Kayser, Rix \&
  Dehnen}{Odenkirchen et~al.}{2009}]{odenkirchen_2009}
Odenkirchen M.,  Grebel E.~K.,  Kayser A.,  Rix H.-W.,    Dehnen W.,  2009, The
  Astronomical Journal, 137, 3378

\bibitem[\protect\citeauthoryear{Odenkirchen, Grebel, Rockosi, Dehnen, Ibata,
  Rix, Stolte, Wolf, Anderson Jr., Bahcall, Brinkmann, Csabai, Hennessy,
  Hindsley, Ivezi{\'c}, Lupton, Munn, Pier, Stoughton \& York}{Odenkirchen
  et~al.}{2001}]{odenkirchen_2001}
Odenkirchen M.,  Grebel E.~K.,  Rockosi C.~M.,  Dehnen W.,  Ibata R.,  Rix
  H.-W.,  Stolte A.,  Wolf C.,  Anderson Jr. J.~E.,  Bahcall N.~A.,  Brinkmann
  J.,  Csabai I.,  Hennessy G.,  Hindsley R.~B.,  Ivezi{\'c} {\v Z}.,  Lupton
  R.~H.,  Munn J.~A.,  Pier J.~R.,  Stoughton C.,    York D.~G.,  2001, The
  Astrophysical Journal Letters, 548, L165

\bibitem[\protect\citeauthoryear{Regnault, Conley, Guy, Sullivan, Cuillandre,
  Astier, Balland, Basa, Carlberg, Fouchez, Hardin, Hook, Howell, Pain, Perrett
  \& Pritchet}{Regnault et~al.}{2009}]{Regnault_2009}
Regnault N.,  Conley A.,  Guy J.,  Sullivan M.,  Cuillandre J.-C.,  Astier P.,
  Balland C.,  Basa S.,  Carlberg R.~G.,  Fouchez D.,  Hardin D.,  Hook I.~M.,
  Howell D.~A.,  Pain R.,  Perrett K.,    Pritchet C.~J.,  2009, Astronomy and
  Astrophysics, 506, 999

\bibitem[\protect\citeauthoryear{Rockosi, Odenkirchen, Grebel, Dehnen,
  Cudworth, Gunn, York, Brinkmann, Hennessy \& Ivezi{\'c}}{Rockosi
  et~al.}{2002}]{rockosi_2002}
Rockosi C.~M.,  Odenkirchen M.,  Grebel E.~K.,  Dehnen W.,  Cudworth K.~M.,
  Gunn J.~E.,  York D.~G.,  Brinkmann J.,  Hennessy G.~S.,    Ivezi{\'c} {\v
  Z}.,  2002, The Astronomical Journal, 124, 349

\bibitem[\protect\citeauthoryear{Sanders \& Binney}{Sanders \&
  Binney}{2013}]{sanders_2013}
Sanders J.~L.,  Binney J.,  2013, Monthly Notices of the Royal Astronomical
  Society, 433, 1826

\bibitem[\protect\citeauthoryear{Schlafly \& Finkbeiner}{Schlafly \&
  Finkbeiner}{2011}]{schlafly_2011}
Schlafly E.~F.,  Finkbeiner D.~P.,  2011, The Astrophysical Journal, 737, 103

\bibitem[\protect\citeauthoryear{Schlafly, Green, Finkbeiner, Juri{\'c}, Rix,
  Martin, Burgett, Chambers, Draper, Hodapp, Kaiser, Kudritzki, Magnier,
  Metcalfe, Morgan, Price, Stubbs, Tonry, Wainscoat \& Waters}{Schlafly
  et~al.}{2014}]{schlafly_2014}
Schlafly E.~F.,  Green G.,  Finkbeiner D.~P.,  Juri{\'c} M.,  Rix H.-W.,
  Martin N.~F.,  Burgett W.~S.,  Chambers K.~C.,  Draper P.~W.,  Hodapp K.~W.,
  Kaiser N.,  Kudritzki R.-P.,  Magnier E.~A.,  Metcalfe N.,  Morgan J.~S.,
  Price P.~A.,  Stubbs C.~W.,  Tonry J.~L.,  Wainscoat R.~J.,    Waters C.,
  2014, The Astrophysical Journal, 789, 15

\bibitem[\protect\citeauthoryear{Schlegel, Finkbeiner \& Davis}{Schlegel
  et~al.}{1998}]{schlegel_1998}
Schlegel D.~J.,  Finkbeiner D.~P.,    Davis M.,  1998, The Astrophysical
  Journal, 500, 525

\bibitem[\protect\citeauthoryear{Smith, Sneden \& Kraft}{Smith
  et~al.}{2002}]{smith_2002}
Smith G.~H.,  Sneden C.,    Kraft R.~P.,  2002, The Astronomical Journal, 123,
  1502

\bibitem[\protect\citeauthoryear{Teuben}{Teuben}{1995}]{teuben_1995}
Teuben P.,  1995 Vol.~77, The {{Stellar Dynamics Toolbox NEMO}}.
p.~398

\bibitem[\protect\citeauthoryear{Varghese, Ibata \& Lewis}{Varghese
  et~al.}{2011}]{varghese_2011}
Varghese A.,  Ibata R.,    Lewis G.~F.,  2011, Monthly Notices of the Royal
  Astronomical Society, 417, 198

\bibitem[\protect\citeauthoryear{Yoon, Johnston \& Hogg}{Yoon
  et~al.}{2011}]{yoon_2011}
Yoon J.~H.,  Johnston K.~V.,    Hogg D.~W.,  2011, The Astrophysical Journal,
  731, 58

\end{thebibliography}

\label{lastpage}

\end{document}